\documentclass[reprint, 
superscriptaddress,
 amsmath,amssymb,
 aps, physrev,
]{revtex4-2}

\usepackage{graphicx}
\usepackage{dcolumn}
\usepackage{bm}
\usepackage{hyperref}
\usepackage{enumitem}
\usepackage{siunitx} 
\usepackage{cleveref}

\begin{document}

\preprint{APS/123-QED}

\title{\textbf{Reducing the weight of low exam scores may raise average grades but does not appear to impact equity gaps} 
}%

\author{Nicholas T. Young}
\email{Corresponding author`s email: nicholas.young@uga.edu}
\affiliation{Department of Physics and Astronomy, University of Georgia, Athens, GA, 30605, USA}
\author{Rebecca L. Matz}
\affiliation{Center for Academic Innovation, University of Michigan, Ann Arbor, MI, 48104, USA}
\author{Eric F. Bell}
\affiliation{Department of Astronomy, University of Michigan, Ann Arbor, MI, 48105, USA}
\author{Caitlin Hayward}
\affiliation{Center for Academic Innovation, University of Michigan, Ann Arbor, MI, 48104, USA}

\date{\today}

\begin{abstract}
Students interpret grades as signals of their strengths, and grades inform students' decisions about majors and courses of study. Differences in grades that are not related to learning can impact this judgment and have real-world impact on course-taking and careers. Existing work has examined how an overemphasis on high-stakes exams can create equity gaps where female students and Black, Hispanic, and Native American students earn lower grades than male students and Asian and white students, respectively. Yet, minimal work has examined how the weighting of individual midterm exam scores can also contribute to equity gaps. In this work, we examine how three midterm exam score aggregation methods for final grades affect equity gaps. We collected midterm exam data from approximately 6,000 students in an introductory physics course over 6 years at a large, research-intensive university. Using this data set, we applied common midterm exam score aggregation methods to determine their impact on aggregated midterm exam grades: dropping the lowest midterm exam score, replacing the lowest midterm exam score with the final exam score if higher, and counting the highest midterm exam score more in the final grade calculation than the lowest midterm exam score. We find that dropping the lowest midterm exam score resulted in the largest increase in final grades, with students with lower grades benefiting the most. However, we find limited evidence that alternative midterm exam aggregation methods could close equity gaps. Implementing the alternative midterm exam aggregation methods examined here may be useful for instructors wanting to raise course grades or give lower-scoring students a boost. However, they do not appear to be effective in reducing equity gaps.
\end{abstract}

\maketitle

\section{Introduction}\label{sec:intro}
Introductory courses have a large impact on undergraduate student retention.
\cite{seymour_talking_2019, castle_systemic_2024}. Final course grades in these courses may be interpreted by students to reflect what they are good at and can inform their choice to remain in or change their major \cite{ost_role_2010, rask_attrition_2010, stinebrickner_major_2014}.

Yet, grades are not assigned equally across different groups of students, even when controlling for prior performance or academic ability. For example, a multi-institutional study found that men tended to earn higher final course grades than women in introductory physics relative to their other courses \cite{matz_patterns_2017}. Other studies found that men tended to earn higher final course grades than women across the physics curriculum \cite{malespina_gender_2022}, and middle- and upper-class, continuing-generation, white, and Asian students tended to earn higher final course grades than students of color, first-generation students, and low-income students \cite{whitcomb_not_2021}. These grade differences affect who earns a science, technology, engineering, and math (STEM) degree, as prior work has found that earning low final course grades early on in STEM courses decreases the chance that a student will graduate with a major in STEM \cite{chen_stem_2013, seymour_talking_2019, thompson_grade_2021}. 

In addition, the effect is not the same for all students. Studies have found that women who failed a key ``weed-out" STEM course were less likely to graduate with a STEM degree than men who failed the same course \cite{sanabria_weeded_2017}. More broadly, men appear to be less sensitive to feedback about their actual course performance and hence, remain in STEM fields despite having lower grades than women who leave STEM \cite{cimpian_understanding_2020}. Further, men have significantly higher levels of self-efficacy than women do, even when men and women's actual performance is similar \cite{marshman_female_2018}. Other work has found that the relationship between earning a lower final course grade in introductory STEM courses and not earning a STEM degree was stronger for students from minoritized races and ethnicities compared to students from majoritized races and ethnicities \cite{hatfield_introductory_2022}.

As physics remains one of the least diverse STEM fields in terms of sex, race, and ethnicity \cite{american_physical_society_statistics_nodate}, various studies have focused on how grading approaches in physics, particularly how final course grades are determined, could be more equitable and contribute to a more equitable education environment. For example, Webb, Paul, and Cheesey found that grading students directly on the 4.0 scale rather than the percent scale both on individual exam items and for the final course grade significantly reduced the number of final course grades less than a C- \cite{webb_relative_2020} and later work found that the same grading system could reduce equity gaps by 20-25\% \cite{paul_percent_2022}. Other work has found that reducing the weight of high-stakes exams in the final course grade would eliminate the male-female grade gap and also reduce equity gaps for underrepresented minority students \cite{simmons_grades_2020}, suggesting that grade component weight, in addition to the grading system, is important for demographic equity. In a similar line of work, Webb and Paul found that allowing retakes of high-stakes exams removed equity gaps for women in introductory physics \cite{webb_attributing_2023} and that the gaps disappeared because the stakes of each exam were reduced, not because women were more likely than men to retake exams or because they used the retake as an opportunity to better learn the material \cite{webb_high_2025}.

Here, we seek to add to these studies by conducting a study of how altering the stakes of individual midterm exams, rather than the overall weighting of midterm exams relative to other graded components, may affect students' final course grades, including equity gaps. We specifically examine the impact of three midterm exam aggregation methods compared to the equal weighting of each midterm exam common in physics courses (e.g \cite{simmons_grades_2020}): dropping the lowest midterm exam grade, replacing the lowest midterm exam grade with the final exam grade if it is higher, and weighting the best midterm exam score more than the worst midterm exam score.

These techniques rest on the assumption that students show significant variations in midterm exam performance over time, such that these alternative aggregation methods could make a meaningful difference in aggregated student final course grades.
Previous work in a single course at a single institution suggests that this is likely the case \cite{suresh_evolution_2023}. Our first research then seeks to confirm that this assumption is met:

\begin{enumerate}
    \item How do student grades vary across midterm exams within an introductory physics course?
\end{enumerate}

Assuming that student midterm exam grades vary over the semester, we then ask:

\begin{enumerate}[resume]
    \item What impact do alternative aggregation methods have on students' overall midterm exam grades as compared to weighting each midterm exam equally?
    \item How do the impacts of the alternative aggregation methods vary based on the students' final grade?
\end{enumerate}

Finally, we then expand the second research question to consider the potential impacts of these alternative aggregation methods on equity gaps:

\begin{enumerate}[resume]
    \item How do the impacts of the alternative aggregation methods vary based on the students’ sex or race?
\end{enumerate}

For Research Questions 2, 3, and 4, we make the critical assumption that student behavior would not change if these alternative aggregation schemes were implemented, an assumption made in similar studies (e.g., \cite{simmons_grades_2020}), and one we discuss further in Sec. \ref{sec:discussion}.

The rest of the paper is organized as follows. In Sec. \ref{sec:framing}, we describe the course deficit model that underlies this work as well as how we are conceptualizing equity to answer Research Question 4. In Sec. \ref{sec:methods}, we describe how we obtained the data for this study, the alternative aggregation methods, and how we created the demographic groups. In Sec. \ref{sec:results}, we show that students exhibit wide variation in midterm exam grades throughout the semester and that while alternative aggregation methods increase grades on average, they do not close equity gaps. In Sec. \ref{sec:discussion}, we provide explicit answers to our research questions and describe the implications of these findings. In Sec.\ref{sec:limit_future}, we discuss the limitations of this study and how future work may address these, including conducting studies in courses that implemented these alternative aggregation methods.

\section{Research Framing}\label{sec:framing}
We use two frameworks to guide our answer to Research Question 4. First, we use the course deficit model to argue that observed equity gaps in a course are the result of the structures of the course rather than an underlying deficiency of one group compared to another \cite{cotner_can_2017, webb_attributing_2023}. That is, course characteristics such as grading policies, content delivery, resources provided, and instructor actions are the cause of equity gaps between different groups of students. Under this paradigm, the solution to closing equity gaps is to alter course structures, rather than to seek to alter student behavior.

Second, because there are multiple definitions of equity \cite{levinson_conceptions_2022, russo-tait_science_2023}, we consider two definitions of equity used within physics education research: equity of parity and equity of fairness \cite{rodriguez_impact_2012}. Equity of parity means that all groups experience the same outcome on average. In the context of this study, if all student groups score, on average, the same on midterm exams after implementing one of the alternative aggregation methods, we would say that method achieved equity of parity because the outcome is the same for all groups, even if there were initial differences. For equity of parity to occur, groups that initially score lower must benefit more from the intervention in order to achieve an equal outcome. This is the common definition of equity used in previous studies when referring to equity gaps (e.g. \cite{whitcomb_not_2021,kost_characterizing_2009, madsen_gender_2013}).

In contrast, equity of fairness requires that all groups receive, on average, the same benefit. If all student groups increase their score by the same amount on average after implementing one of the alternative midterm exam aggregation methods, we would say that method achieved equity of fairness. Because all groups receive the same benefit, any existing performance differences between groups before the intervention are maintained after the intervention.

\section{Methodology}\label{sec:methods}

\subsection{Data collection}

\begin{table}[]
    \centering
    \caption{The four assessment styles used in the course of interest and the number of students included in the analysis for research questions 2 through 4}
    \begin{tabular}{p{0.19\linewidth}p{0.1\linewidth}p{0.65\linewidth}} \hline
    Assessment Style & N & Description\\ \hline
    3E + F & 2,290 & Physics I sections between Fall 2017 and Fall 2019 had three in-person midterm exams and one final exam.\\
    4E + F & 1,274 & Physics I sections between Fall 2022 and Fall 2023 had four in-person midterm exams and a final exam, with the final exam scores recorded in ECoach. \\ 
    4E & 750 & Physics I sections in Fall 2021 and Winter 2022 had four in-person midterm exams and a final exam but the instructors did not include final exam grades in ECoach. \\
    5E+ F & 767 & Physics I sections in Fall 2020 and Winter 2021 had five online midterm exams and one final exam.  \\\hline
    \end{tabular}
    \label{tab:assessment_styles}
\end{table}

Data for this study come from ECoach users enrolled in a first-semester, calculus-based introductory physics course (referred to as Physics I) between Fall 2017 and Fall 2023, excluding the Winter 2020 semester when courses and course grading policies were significantly impacted by the COVID-19 pandemic. ECoach is a free, personalized, web-based coaching tool offered to students in introductory STEM courses (including Physics I) at the University of Michigan \cite{huberth_computer-tailored_2015, matz_analyzing_2021}. Student use is optional in the course under study, but instructors typically incentivize participation by offering a small amount of extra credit, equivalent to an in-class assignment or two, for students who choose to use ECoach.

ECoach provides students with tips on how to approach their course, personalized feedback on their performance, and tools to help them succeed, such as task lists, grade calculators, and midterm exam study playbooks. In order to provide personalized feedback and for the grade calculator to provide accurate estimates, assignment grades for participating students are pulled into the platform from the learning management system, allowing the ECoach gradebook used in this study to act as a duplicate of instructors' official gradebook. Midterm and final exam scores in the gradebook included both the total points possible on each exam and the number of points the student earned. The total possible points on the midterm exams ranged from 12 points to 20 points (median 15 points) and the total possible points on the final exams ranged from 25 points to 26 points (median 25 points). Students were only able to earn a whole number of points on each exam. 

During the time period of study and excluding Winter 2020, there were 60 Physics I sections offered. Of these 60 sections, 30 instructors offered ECoach to their students while the remaining 30 sections did not offer ECoach. In the 30 sections that did offer ECoach, there were 6,082 student enrollments and 5,621 of those students enrolled in ECoach, resulting in a usage rate of approximately 92\%. Repeat enrollments are included in this count.

Data for this study also came from a simplified version of the university's student records data warehouse \cite{lonn_rearchitecting_2019}. Through the data warehouse, we obtained students' final grades in Physics I, their sex, and their race and ethnicity.

Due to the variety of assessment styles used by Physics I instructors throughout the study period and recorded in ECoach, we created four separate data sets to correspond to the distinct assessment styles. For the remainder of the paper, we refer to each of these assessment styles by the number of midterm exams offered (E) and whether there was a final exam (F). These styles and a description are shown in Table \ref{tab:assessment_styles}.

While the specific weighting of midterm exams and the final varied in each section, the median weighting of the midterm exams was 36\% of the final grade (range 35\% to 45\%) and the median weighting of the final exam was 16\% of the final grade (range 15\% to 19\%). The weight of each individual midterm exam was equal to the total weight divided by the number of exams. In each section, high-stakes exams accounted for at least half of each student's final grade, with homework, in-class participation, and pre-lecture activities accounting for the remaining portion of students' final grades.

For each of these four data sets, we apply the same analysis, which is detailed in the next section.

\subsection{Data handling and analysis}

\subsubsection{Grade variations}

\begin{table}
    \centering
    \caption{Grade ranges for Physics I, ignoring pluses and minuses}
    \begin{tabular}{cc} \hline
         Grade& Percentage range\\ \hline
         A& 85-100\\
         B& 75-84.99\\
         C& 60-74.99\\
         D& 45-59.99\\
         E& 0-44.99\\ \hline
    \end{tabular}
    \label{tab:grading_table}
\end{table}

To answer Research Question 1, all midterm exam scores were converted to percentages and then to letter grade using the grading scale provided by instructors in ECoach. Although the courses used a plus-minus grading system, we used whole letter grades to reduce the number of potential grades from 13 to 5. The percentage-to-grade conversation is shown in Table \ref{tab:grading_table}. We do this because as the exam grades are whole numbers, many of the letter grades are impossible for students to earn on an individual assignment, and there is often only one possible midterm exam score for a student to earn each letter grade. For example, for midterm exams graded out of 15 points, it would be impossible for the student to earn a D+, B-, or B+ on that midterm exam based on the course grading scheme, and there is only a single score each that results in a D-, D, C-, C, C+, B, or A-.

For each midterm exam and its subsequent midterm exam (e.g., exam 2 and exam 3), we calculated the fraction of students who earned each letter grade on the subsequent midterm exam given the letter grade on the midterm exam of interest, essentially forming a ``transition matrix'' from one midterm exam grade to the next (similar to Morris et al. \cite{morris_transition_2017}). We then aggregated these fractions into the fraction of students who earned the same letter grade, earned a higher letter grade on the subsequent midterm exam, and earned a lower letter grade on the subsequent midterm exam. Students with missing midterm exam data were removed pairwise, meaning that a student could be missing a score on exam 3 but still be included in the exam 1 to exam 2 analysis.

Second, because the grade students earn on a midterm exam depends on both their performance and the underlying difficulty of the midterm exam, we also normalized each midterm exam score, using the mean and standard deviation of each individual midterm exam to normalize. We grouped these normalized midterm exam scores into quintiles, picking five groups as there are five potential letter grades. These quintiles allow us to compare how students perform relative to their peers across midterm exams rather than only examine how their letter grades vary across midterm exams.

Next, we repeated the analysis with the quintiles. We calculated the fraction of students in each quintile on the subsequent midterm exam given their quintile on the midterm exam of interest. We then aggregated these fractions into the fraction of students who fell in the same quintile, fell into a higher quintile on the subsequent midterm exam, and fell into a lower quintile on the subsequent midterm exam.

Finally, we calculated the Kendall correlation \cite{kendall_new_1938} between midterm exam letter grades on previous and subsequent midterm exams and between the quintiles for the previous and subsequent midterm exams. We chose to use the Kendall correlation because the data are ordinal and not continuous and because there are only five possible outcomes, multiple ties were present. We calculated the Kendall correlation as $\tau_b$ instead of the default $\tau_a$ due to the presence of a large number of ties \cite{kendall_treatment_1945}.

As the Kendall correlation is a measure of monotonicity, a large correlation means that students who earn a low grade on an midterm exam tend to earn a similarly low grade on the subsequent midterm exam, and students who earn a high grade on an midterm exam tend to earn a similarly high grade on the subsequent midterm exam.

For completeness, we also include Pearson correlation coefficients with the numerical midterm exam scores in the supplemental material.

\subsubsection{Alternative aggregation methods}
To answer Research Questions 2 through 4, we conducted a thought experiment around how three alternative midterm exam score aggregation methods could affect student grades. As we needed all midterm exam grades (including the final as relevant), students with any missing midterm or final exam grades were excluded from this analysis. The number of students remaining for each assessment style are shown in Table \ref{tab:assessment_styles}.

In this paper, we examined three potential alternative aggregation methods based on knowledge of what introductory physics instructors at other institutions have tried in their courses. These methods, how they were implemented, and why instructors may use them, are described below:

\begin{itemize}
    \item \textbf{Drop the lowest midterm exam score:} The student's lowest midterm exam score is dropped and the total weight of the midterm exams is split between the remaining midterm exams. For example, for a section with three midterm exams that in total contribute 36\% to the final grade, the two best midterm exam scores would each contribute 18\% of a student's final grade. Such an approach allows a student to earn an abnormally low grade without lowering their overall grade in the course, whatever the reason (e.g. not understanding the course material or personal factors). However, in preliminary discussion of this manuscript with instructors, some worried that students would not study for or take the last midterm exam if they were already satisfied with their grades on the previous midterm exams because the last midterm exam score would be dropped regardless.
    \item \textbf{Replace the lowest midterm exam score with the final:} If a student scores better on the final exam than they did on their lowest midterm exam score, the final exam score replaces the lowest midterm exam score. Using the example in the previous bullet point, if the student scored higher on the final exam, the final exam would take the place of the lowest midterm exam and count for 12\% of the student's final grade (36\%/3) \textit{in addition} to the regular weighting of the final exam. Such an approach rewards students for growth over time by allowing them to replace a low score with a higher score on the (cumulative) final exam.
    \item \textbf{Weigh the best midterm exam score more than the lowest midterm exam score:} The student's best score counts for 50\% more and the student's lowest score counts for 50\% less than the remaining midterm exams. For the same example section, the best midterm exam score would count for 18\% of the final grade, the middle score would count for 12\% of the grade, and the lowest score would count for 6\% of the final grade. Such an approach combines the benefits of the previous two approaches, rewarding growth and limiting the impact of an abnormally low midterm exam score by weighting the highest midterm exam score the most and weighting the lowest midterm exam score the least. In addition, because all midterm exams still have a non-zero weight in the final course grade calculation, students cannot skip the last midterm exam without consequence, addressing an instructor concern about dropping the lowest midterm exam score.
\end{itemize}

The grading scheme used in the course (referred to here as \textit{default}) divided the total weight of the midterm exams by the number of midterm exams to get the weight of each midterm exam. Again using the example three midterm exam course worth 36\% of the final grade, each midterm exam would have a weight of 12\%.

For each student, we calculated their default aggregated midterm exam grade as well as their aggregated midterm exam grade under each modification. To calculate the default aggregated grade, all midterm exam scores as percentages used in the calculation were averaged. Because each midterm exam was worth the same number of points, this is equivalent to adding up the total points a student earned on midterm exams and dividing by the total possible points on the midterm exams.

To determine whether any of the three alternatives increased aggregated midterm exam scores relative to the default method, we conducted a \textit{t}-test between the aggregated midterm exam grade from the default midterm exam scoring method and the aggregated midterm exam grade from each of the three alternatives. To determine the practical significance of the differences, we also calculated the effect size using Cohen's \textit{d}. We use the standard $\alpha=0.05$ as our cut-off for statistical significance and $d\geq0.2$ to be the minimum effect size for a non-negligible effect \cite{cohen_statistical_1988}.

\subsubsection{Final course grades and demographics}

\begin{table*}
    \centering
    \caption{Number of students in each group and missing for each time period in the study.}
    \begin{tabular}{p{0.07\linewidth} p{0.07\linewidth}p{0.09\linewidth}p{0.09\linewidth} p{0.09\linewidth}p{0.11\linewidth}p{0.09\linewidth} p{0.07\linewidth}p{0.11\linewidth}p{0.09\linewidth}} \hline
         Period&  Male&  Female&  Sex missing & A/w&  B/H/M/N&  Race missing & A& B, C, D, E & Final grade missing\\ \hline
         3E+F&  1,559&  723& 8 &  1780&  374& 136 & 599& 1,676 & 15\\
         4E+F&  791&  479& 4 & 940&  257& 77 & 371& 894 & 9\\
         4E&  472&  276& 2 & 576&  128& 46 &  240& 506 & 4\\
         5E+F&  513&  251& 3 & 587&  127& 53 & 162& 603 & 2\\ \hline
    \end{tabular}
    \label{tab:demographics}
\end{table*}

To answer Research Questions 3 and 4, we used final course grades and demographics from the university's data warehouse.

Grades are reported as letter grades with the plus-minus system. For the purposes of this paper, we grouped students who earned an A-, A, or A+ in Physics I overall as high-grade earners and all other students as lower-grade earners, similar to prior work \cite{young_exploring_2025}. We do this because when physics instructors refer to students who earn high grades in physics, they are generally referring to students who earn ``A''s rather than, for example, students who earn ``B"s even though those grades are still above average grades.

The university's data warehouse only collects data about sex and not gender, so results are split by sex. Sex is reported as \texttt{male}, \texttt{female}, or \texttt{unknown} in the data warehouse. For the purposes of this paper, we assume students who selected male are male and students who selected female are female, ignoring that sex is not binary \cite{westbrook_new_2015,dignazio_data_2020}. We treat students marked as unknown as missing data. Our analysis then only compares students categorized as male with students categorized as female.

For race and ethnicity, students can select multiple options to describe their race and ethnicity. Students could select \texttt{Asian, Black, Hispanic, Native American/Hawaiian}, or \texttt{white} as possible options. We grouped Asian and white students into one group (called \texttt{A/w}) and all other students into a second group, following how similar studies have created binary racial groups \cite{simmons_grades_2020, suresh_evolution_2023}. Students who selected multiple racial or ethnic groups, regardless of which racial or ethnic groups were selected, were included in this second group. Because this group includes students who selected Black, Hispanic, Native American, and multiple categories, we called this group \texttt{B/H/M/N}. We acknowledge that aggregating groups of students can hide differences between groups \cite{shafer_impact_2021}.

As we are interested in differential effects, we computed the difference between the aggregated midterm exam score from each grading modification and the aggregated midterm exam score from the default grading scheme for each student, equivalent to the grade increase as a result of each modification. We then compared the average grade difference between each of the alternative midterm exam score aggregation methods and the default grading scheme for the two groups of students. To determine whether the differences were statistically significant, we performed t-tests. We also calculated the effect size for each difference, again using Cohen's d. We did this for students divided by sex, then race, and then by final grade. For each analysis, we removed any student with missing sex, race, or final grade information only for that specific analysis. For example, a student who did not report their race but did report their sex and final grade would only be excluded from the race analysis. The number of students in each group for each analysis is shown in Table \ref{tab:demographics}.

In terms of our research framing, if we find that the grade difference between, for example, male and female students, is not statistically different between one of the alternative aggregation schemes and the default method, we would have evidence of equity of parity. If we do not find a statistically or practically significant difference between the grade increases by each of the alternative aggregation methods for each group, we would have evidence of equity of fairness.

\section{Results}\label{sec:results}

\subsection{Grade variations}

\begin{figure*}
    \centering
    \includegraphics[width=0.9\linewidth]{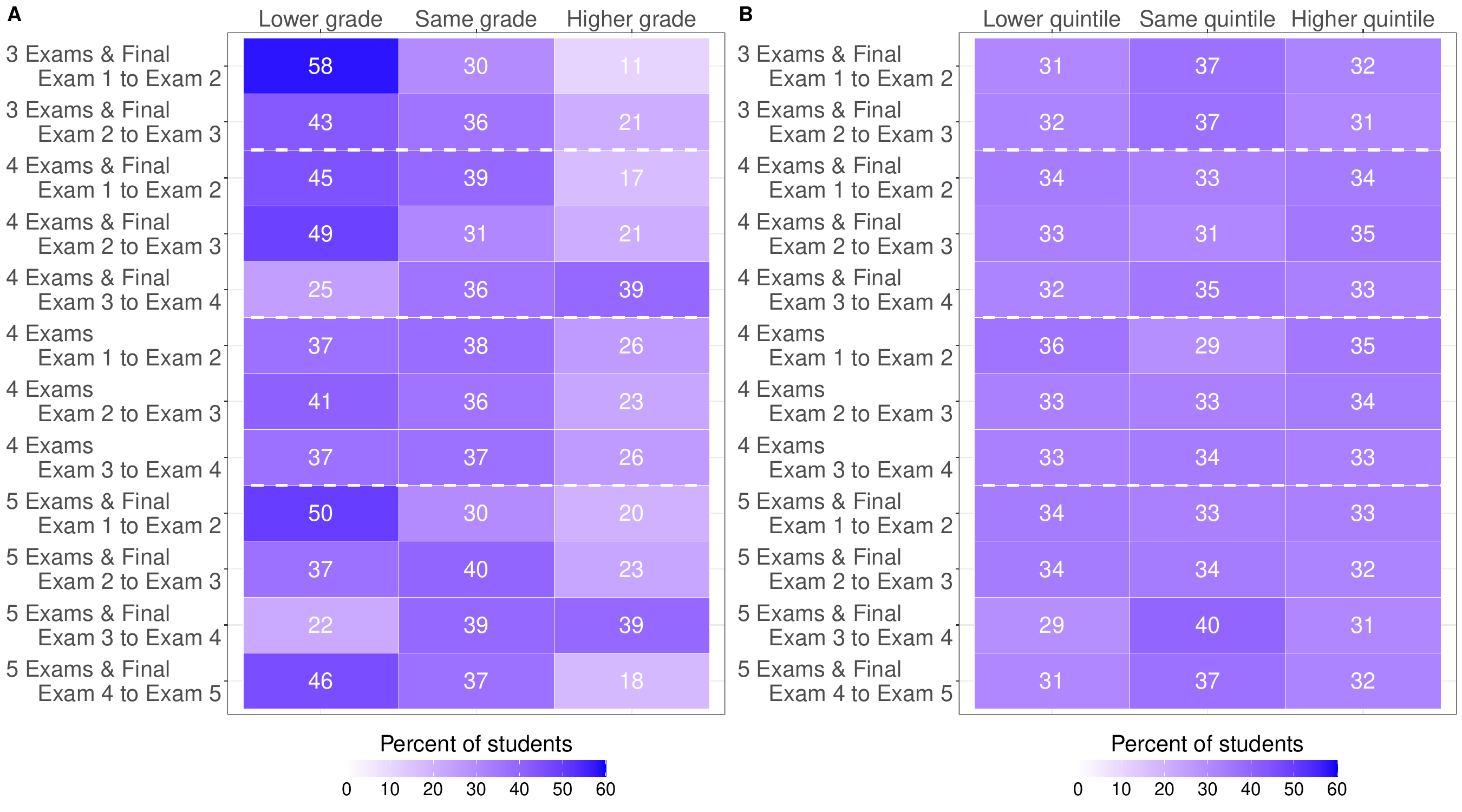}
    \caption{Percent of students earning a lower letter grade, the same letter grade, or a higher letter grade on each subsequent midterm exam (plot A) and students in a lower quintile, the same quintile, and a higher quintile on each subsequent midterm exam (plot B). Numbers in the boxes are the percentage of students. Rows may not add up to exactly 100\% due to rounding. Students earning a lower letter grade on the subsequent midterm exam was the most common grade transition, and students being in the same quintile on the subsequent midterm exam was the most common quintile transition.}
    \label{fig:transitions}
\end{figure*}

Overall, we find 43\% of students earn a lower letter grade on their subsequent midterm exam, 22\% earn a higher letter grade on their subsequent midterm exam, and 35\% earn the same letter grade on their subsequent midterm exam.

Breaking the results out by each midterm exam, Fig. \ref{fig:transitions}A, we find that in a majority of the section types and midterm exam transitions, the most common outcome is still that students earn a lower grade on the subsequent midterm exam. This result occurred in 7 of the 12 midterm exam transitions. In addition, earning a lower grade on the subsequent midterm exam and earning the same grade on the subsequent midterm exam were tied for the most common result in 2 of the 12 midterm exam transitions. Of the remaining 3 midterm exam transitions, the most common result was students earning the same grade on the subsequent midterm exam in one case, earning a higher grade on the subsequent exam in the second case, and earning the same grade or higher grade.

We find that earning the same grade and earning a higher grade were the next most likely transition to occur, with this occurring in 2 of the 12 midterm exam transitions each.

Examining quintiles overall, we find that 35\% of students place in the same quintile on the subsequent midterm exam, 33\% place in a higher quintile on the subsequent midterm exam, and 32\% place in a lower quintile on the subsequent midterm exam.

Breaking the results out by each midterm exam again, Fig. \ref{fig:transitions}B, we still find that remaining in the same quintile on the subsequent midterm exam is the most common result, occurring in 6 of the 12 transitions. We find that having approximately the same percentage of students move up a quintile as moving down a quintile as the second most common result, occurring in 3 of the 12 transitions. We also found a more uniform distribution of outcomes when examining quintiles compared to the letter grades with each transition having at least 29\% of the students and no more than 40\% of the students for the quintiles while at least one of the possible outcomes (lower, same, or higher) had fewer than 29\% of students for all 12 transitions when looking at letter grades.

\begin{figure*}
    \centering
    \includegraphics[width=0.9\linewidth]{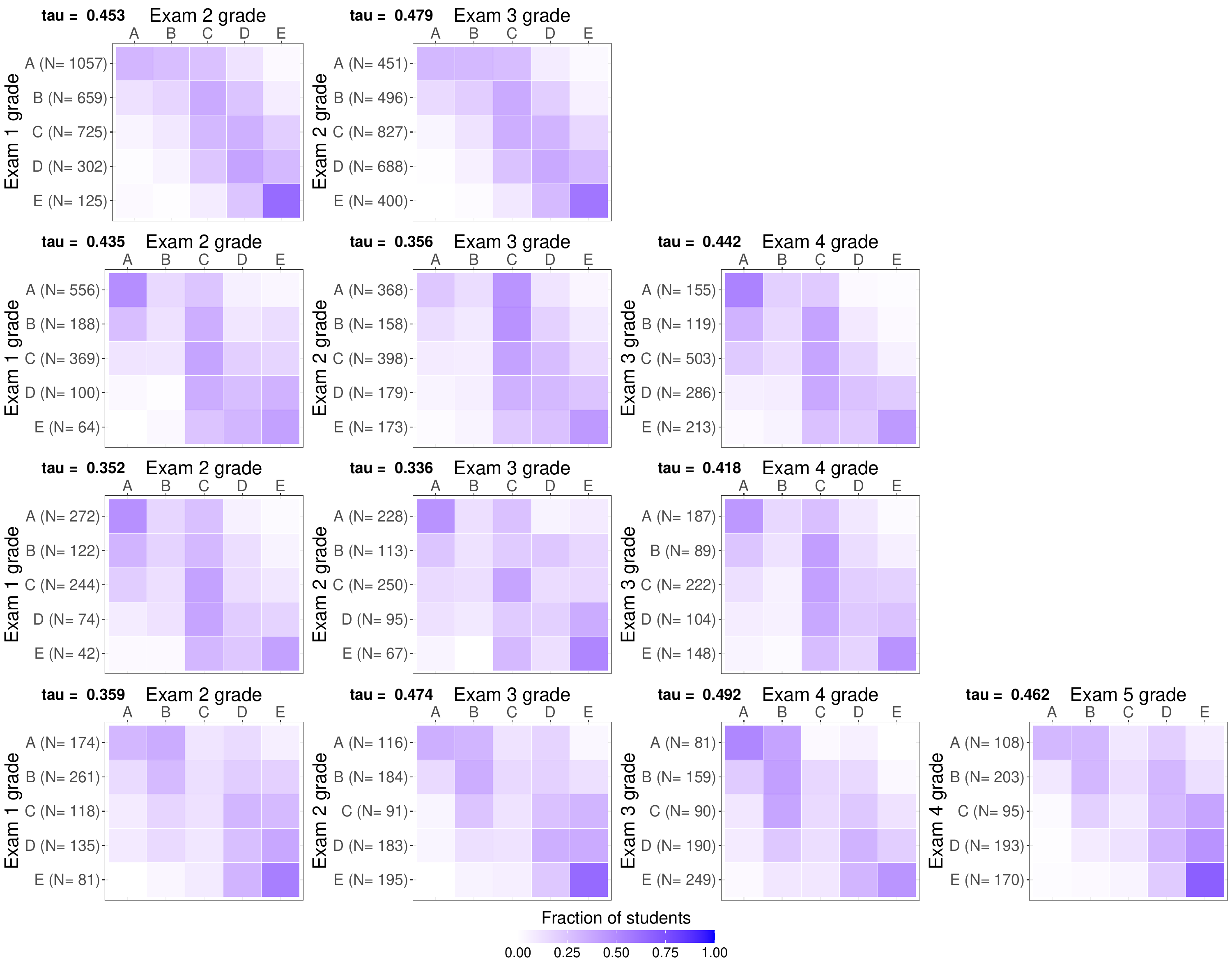}
    \caption{Fraction of students transitioning from one midterm exam grade to the other letter grades on the subsequent midterm exam and the Kendall correlation. From top to bottom, plots show the three midterm exams and a final, the four midterm exam and a final, the four midterm exams no final, and five midterm exams and a final courses. In general, we find that students who earn ``A''s or ``E''s on one midterm exam tend to earn those grades on subsequent midterm exams and most students tend to earn a letter grade on the subsequent midterm exam within a step of their grade on the current midterm exam.}
    \label{fig:expanded_transitions}
\end{figure*}

\begin{figure*}
    \centering
    \includegraphics[width=0.9\linewidth]{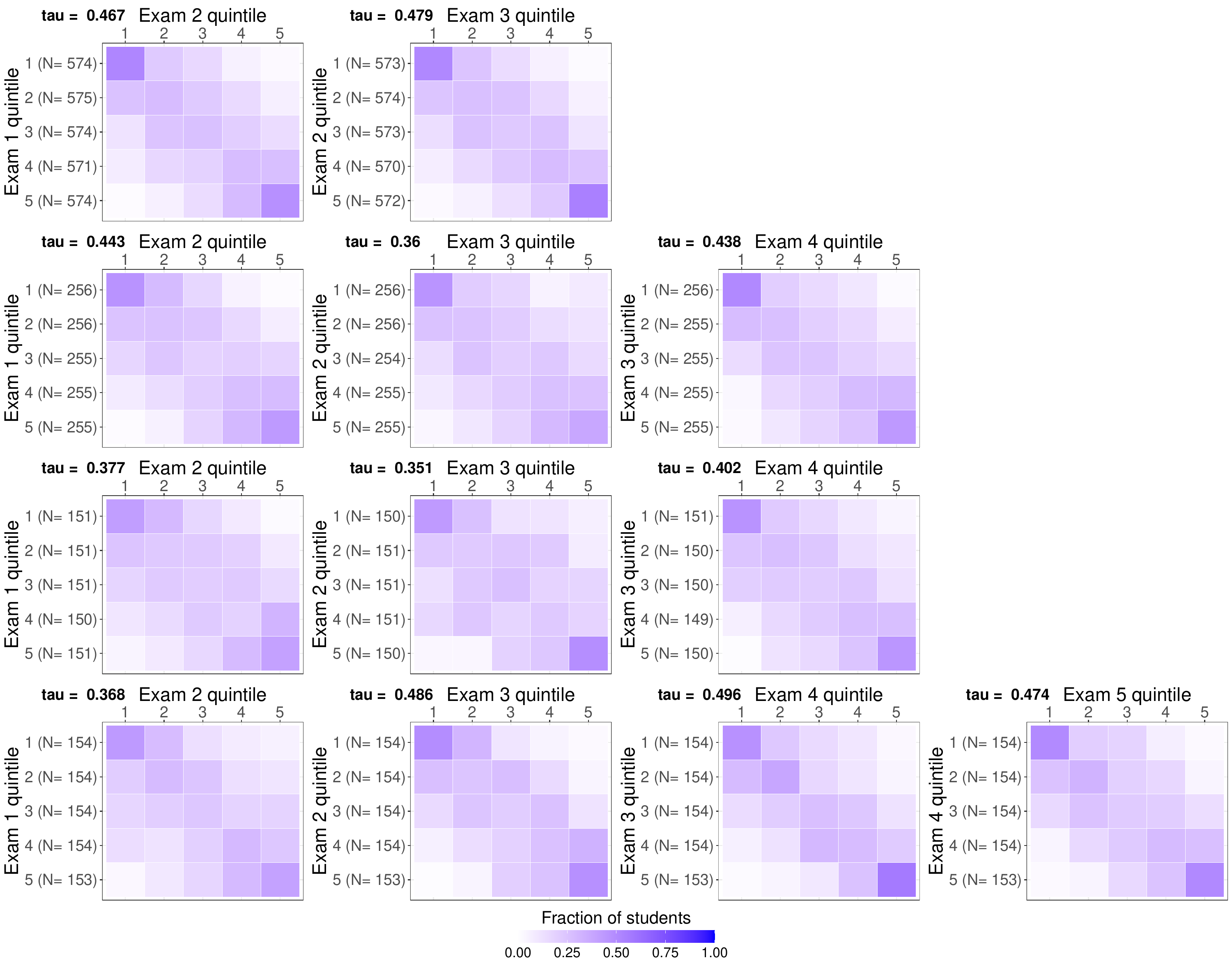}
    \caption{Fraction of students transitioning from one midterm exam quintile to the other quintiles on the subsequent midterm exam and the Kendall correlation. From top to bottom, plots show the three midterm exams and a final, the four midterm exam and a final, the four midterm exams no final, and five midterm exams and a final courses. In general, we find that students in the top or bottom quintile on one midterm exam tend to stay in those quintiles on subsequent midterm exams and most students tend to score within a quintile on the subsequent midterm exam as they do on the current midterm exam.}
    \label{fig:expanded_transitions_quintile}
\end{figure*}

To gain better understanding of these transitions, we then disaggregated the results by the specific letter grade earned or quintile in on the preceding midterm exam. The results are shown in Figs. \ref{fig:expanded_transitions} and \ref{fig:expanded_transitions_quintile}.

For the letter grades, we find that students at the top and bottom of the grading scale (A and E grades) tend to also earn those grades on the subsequent midterm exams. We find very few students who move from one end of the grading scale (A or E) to the other (E or A).

For students who do earn a different grade on the subsequent midterm exam, we find that most of these students tend to earn a letter grade within one step of their prior midterm exam grade. Examining the Kendall correlations, we find that they range from 0.336 to 0.492 suggesting a small to moderate correlation between one midterm exam letter grade and the subsequent letter grade.

We find a similar result for the quintile plots. In general, students in the top or bottom quintiles tend to stay in the top or bottom quintile on subsequent midterm exams. If a student does switch between quintiles, they also tend to not move more than one quintile from their current position. Examining the Kendall correlations, we find that they range from 0.360 to 0.496, again suggesting a small to moderate correlation between one midterm exam quintile and the subsequent midterm exam quintile.

\subsection{Alternative aggregation methods}

\begin{figure}
    \centering
    \includegraphics[width=0.9\linewidth]{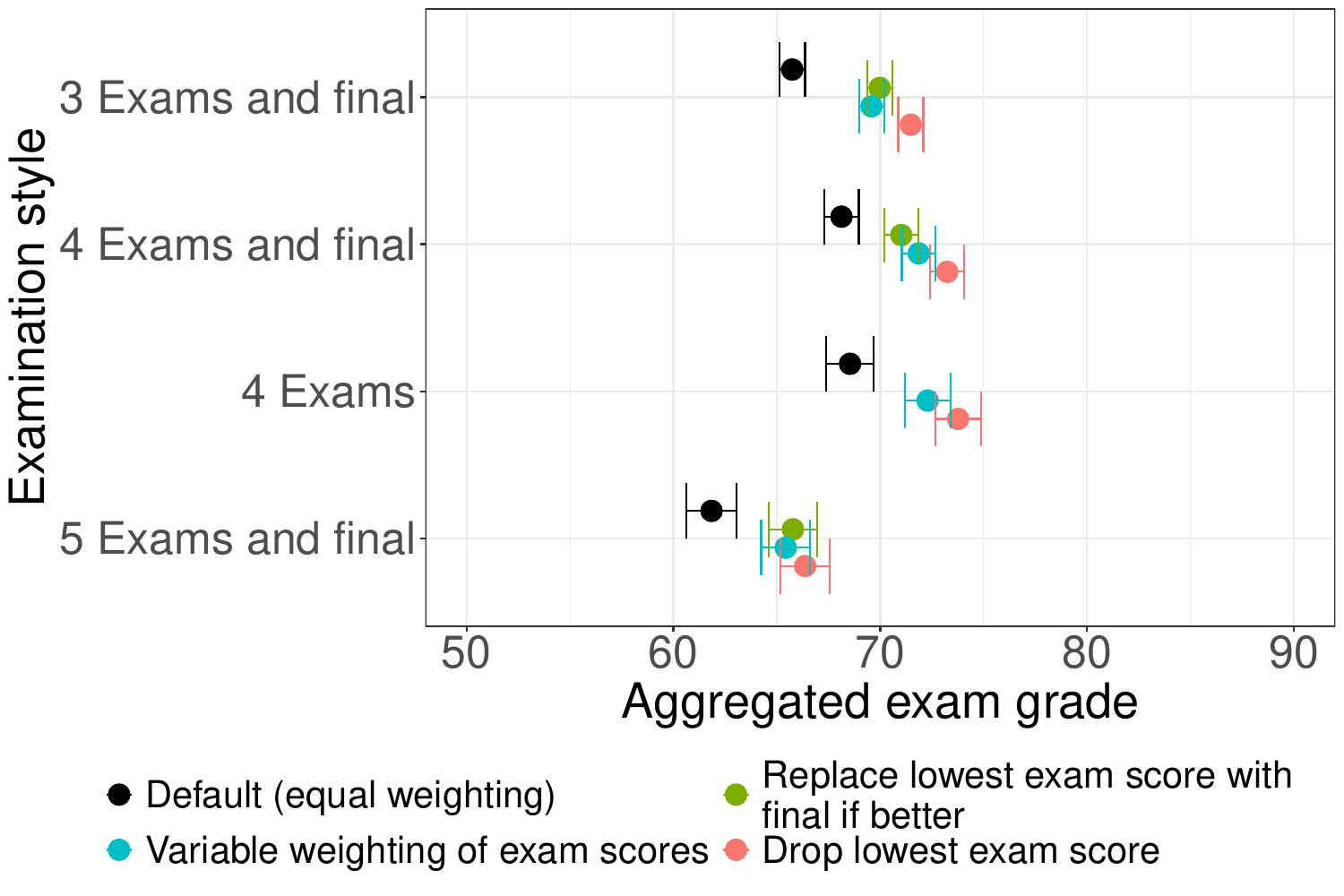}
    \caption{Average aggregated midterm exam grade for the default equal weighting and each of the three alternative methods for the four assessment styles. All alternative methods increase grades with dropping the lowest midterm exam score resulting in the largest increase in grades.}
    \label{fig:overall_grades}
\end{figure}

Results from the alternative midterm exam aggregation methods are shown in Fig. \ref{fig:overall_grades}. We find that dropping the lowest midterm exam score resulted in the largest increase in grades under all four assessment styles, with a minimum increase of 4.5 percentage points. In course sections with a final exam, we find that replacing the lowest midterm exam score with the final exam score if higher and variably weighting the midterm exam scores resulted in a similar increase in grades. 

We find that each of the aggregated grades are statistically different from the default aggregated midterm exam grade with $p<0.001$. We also find that implementing each of the alternative aggregation methods would have a small effect size, with the exception of replacing the lowest midterm exam with the final in the 4 midterm exam and a final period, which had no effect. These effect sizes are shown in Table \ref{tab:effectsizes}.

\begin{table*}
    \centering
    \caption{Effect sizes of the three alternative midterm exam aggregation methods compared to the default equal weighting. Positive values mean that the alternative midterm exam aggregation method resulted in higher midterm exam scores than the equal weighting method.}
    \begin{tabular}{p{.26 \linewidth}p{.23\linewidth}p{.23\linewidth}p{.23\linewidth}}
        \hline
        Format & Drop lowest vs default & Replace with final vs default & Variable weighting vs default \\ \hline
        3 midterm exams \& final & 0.38 & 0.28 & 0.26 \\ 
        4 midterm exams and final & 0.34 & 0.19 & 0.24 \\ 
        4 midterm exams & 0.33 & NA & 0.24 \\ 
        5 midterm exams \& final & 0.27 & 0.23 & 0.21 \\ \hline
    \end{tabular}
    \label{tab:effectsizes}
\end{table*}

\subsection{Final grades}

\begin{figure*}
    \centering
    \includegraphics[width=0.5\linewidth]{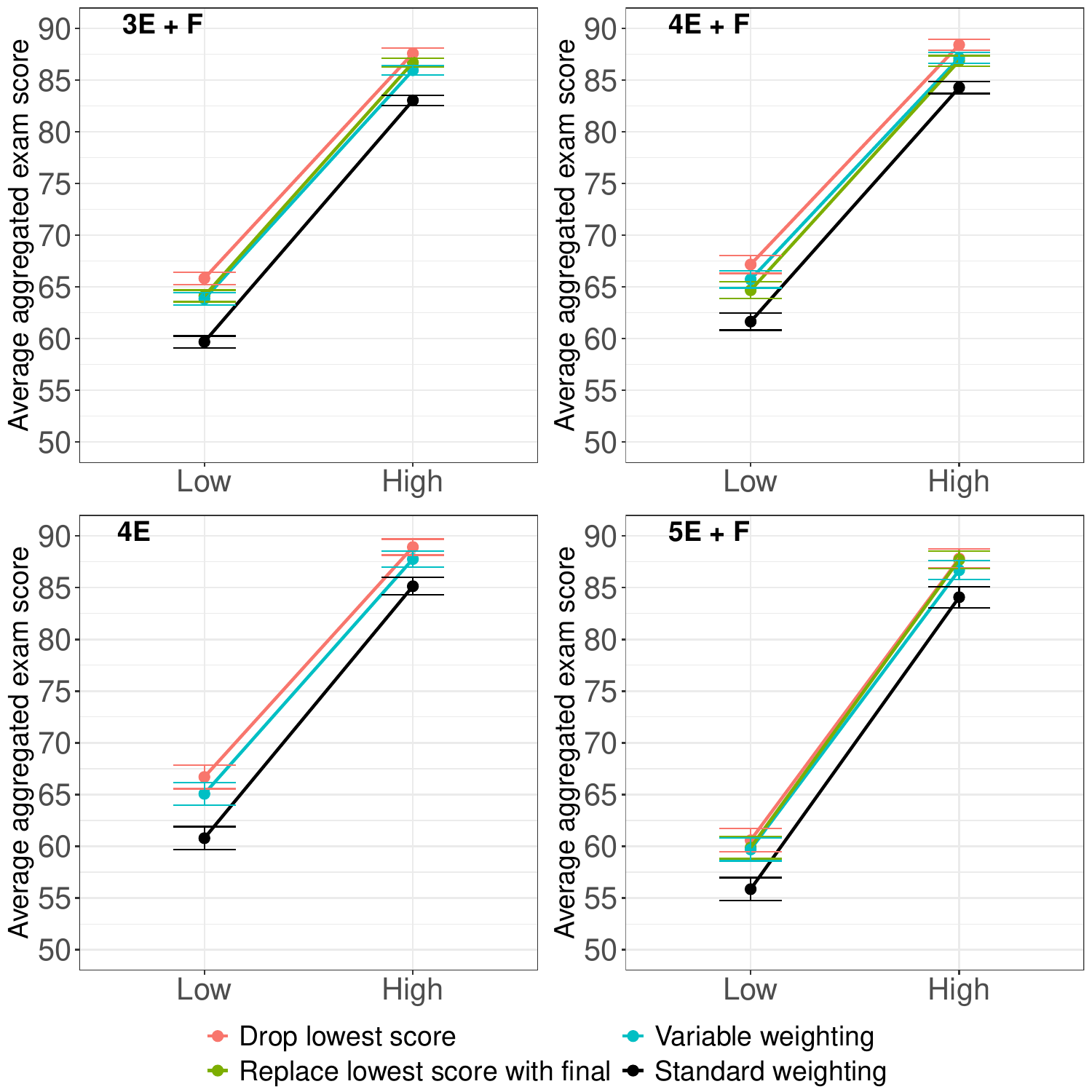}
    \caption{Average aggregated midterm exam scores for students who earned an ``A'' in Physics I (High) and students who earned a ``B'', ``C'', ``D'', or ``E'' in Physics I (low) according to the aggregation method used and the assessment style. Note that the vertical axis does not start at zero. In all courses, ``A'' students outperformed the other students but students who earned a ``B'', ``C'', ``D'', or ``E'' generally received a larger benefit.}
    \label{fig:grade_overall}
\end{figure*}

\begin{table*}
    \centering
    \caption{Difference in increases in aggregated midterm exam scores compared to the default method between students who earned an ``A'' in Physics I and students who earned a ``B'', ``C'', ``D'', or ``E'' in Physics I. Positive values mean that students who earned an ``A'' received a larger benefit and the grade gap increases. The top value represents the aggregated midterm exam score difference (reported in percentage points) with $p<.001$ represented as ***, $p<0.01$ as **, and $p<0.05$ as *. The bottom value represents the effect size.}
    \begin{tabular}{p{.15\linewidth}S[table-format=-1.2]S[table-format=1.2]S[table-format=-1.2]}    \hline
    Format & {Drop lowest} & {Replace with final} & {Variable weighting} \\ \hline
    3 midterm exams \& final & -1.59*** & -0.79***  & -1.24*** \\
     & -0.53 & -0.21  & -0.66 \\ 
    4 midterm exams and final & -1.40*** & -0.45** & -1.23*** \\
     & -0.59  & -0.16 & -0.82 \\
    4 midterm exams & -2.13*** & NA & -1.66** \\
     & -0.80 & NA & -1.05 \\
     5 midterm exams \& final & -0.99***  & -0.39  & -1.22*** \\
     & -0.47 & -0.14 & -0.89 \\ \hline
    \end{tabular}
    \label{tab:grades}
\end{table*}

When examining the results based on the final grade earned in Physics I (Fig. \ref{fig:grade_overall}), we still find that dropping the lowest midterm exam scores results in the largest aggregated midterm exam score increase and that the students who earned ``A''s in Physics I always outperform the other students on average.

Examining the differential impacts, we now find that students who earned a Physics I grade of ``B'' or lower  differentially benefited from these alternative aggregation methods. We find a statistically significant difference of at least 0.45 percentage points and as much as 2.13 percentage points in 10 of the 11 aggregation-method-and-assessment-style pairs. Further, 9 of these 10 aggregation-method-and assessment-style pairs also had at least a small effect size, and 4 of them had a large effect size. The aggregation-method-and assessment-style pairs that were not statistically significant or did not have at least a small effect size were found under the replace the lowest midterm exam score with the final. These are shown in Table \ref{tab:grades}.

\subsection{Demographics}

\begin{table*}
    \centering
    \caption{Difference in increases in aggregated midterm exam scores compared to the default method between male and female students (A) and between Asian and white students and Black, Hispanic, Multiracial, and Native American students (B). Positive values mean that males (A) or Asian and white students (B) received a larger benefit and the grade gap increases. The top value represents the aggregated midterm exam score difference (reported in percentage points) with $p<.001$ represented as ***, $p<0.01$ as **, and $p<0.05$ as *. The bottom value represents the effect size, all of which are negligible.}
    \begin{minipage}{0.48\textwidth}
        \centering
        \textbf{Sex} \\[6pt]
        \begin{tabular}{p{.17\linewidth}S[table-format=-1.2]S[table-format=1.2]S[table-format=-1.2]}
        \hline
        Format & {Drop lowest} & {Replace with final} & {Variable weighting} \\ \hline
        3 midterm exams \& final & -0.18 & 0.11 & -0.14 \\ 
        & -0.06 & 0.03 & -0.07 \\
        4 midterm exams \& final & -0.45** & -0.15 & -0.17 \\ 
        & -0.18 & -0.05 & -0.10 \\
        4 midterm exams & 0.51* & NA & 0.33* \\ 
        & 0.18 & NA & 0.18 \\ 
        5 midterm exams \& final & 0.02 & 0.13 & 0.04 \\ 
        & 0.01 & 0.04 & 0.03 \\\hline
        
    \end{tabular}
    \end{minipage}
    \hfill
    \begin{minipage}{0.48\textwidth}
        \centering
        \textbf{Race} \\[6pt]
    \begin{tabular}{p{.17\linewidth}S[table-format=-1.2]S[table-format=1.2]S[table-format=-1.2]}    \hline
    Format & {Drop lowest} & {Replace with final} & {Variable weighting} \\ \hline
    3 midterm exams \& final & -0.04 & -0.18 & -0.03 \\
     & -0.01 & -0.04 & -0.02 \\ 
    4 midterm exams \& final & -0.33 & 0.07 & -0.19 \\
     & -0.13 & 0.03 & -0.11 \\
    4 midterm exams & -0.30 & NA & -0.18 \\
     & -0.10 & NA & -0.10 \\ 
    5 midterm exams \& final & 0.28 & 0.53 & 0.02 \\ 
     & 0.13 & 0.18 & 0.01 \\ 
 \hline
    \end{tabular}
    \end{minipage}
    \label{tab:demographic_results}
\end{table*}

\begin{figure}
    \centering
    \includegraphics[width=0.9\linewidth]{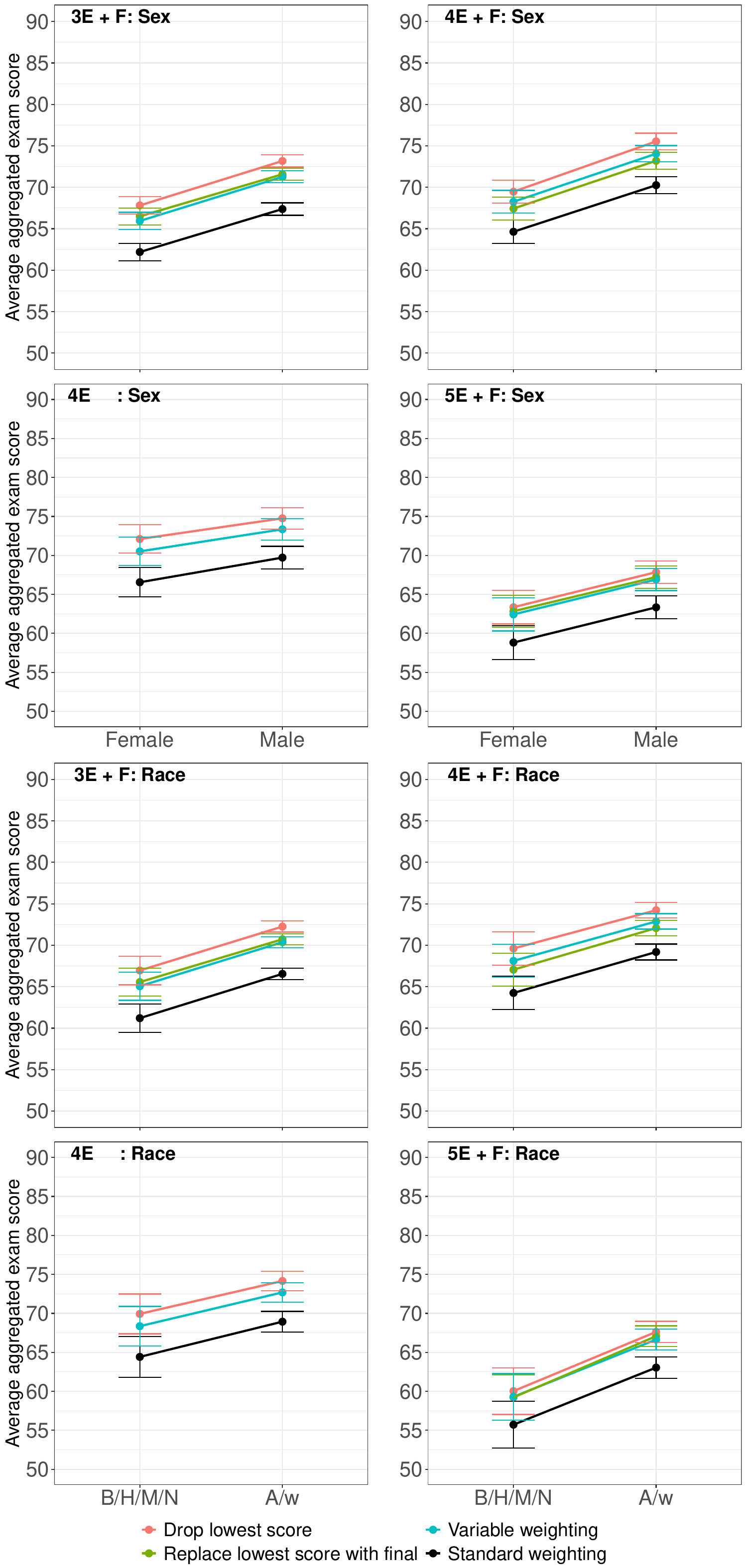}
    \caption{Average aggregated midterm exam scores by sex, and race according to the aggregation method used and the assessment style. Note that the vertical axis does not start at zero. In all course sections, male students outperformed female students and Asian and white students outperform Black, Hispanic, Multiracial, and Native American students. We found limited evidence of differential effects based on sex or race however.}
    \label{fig:demographics_overall}
\end{figure}

Starting by comparing male and female students, we find similar results to the last section in that dropping the lowest midterm exam score continues to result in the largest grade increase (Fig. \ref{fig:demographics_overall}).

Across all assessment styles, we find that on average, male students earned statistically higher midterm exam grades than female students. We find this result to be true regardless of whether we use the default equal-weighting aggregation scheme or whether we use one of the three alternative aggregation schemes.

The differences in increases in grades between male and female students are shown in Table \ref{tab:demographic_results}. We find these differences to be relatively small with the largest difference being about half a percentage point. We find that 8 of the 11 aggregation-method-and assessment-style pairs do not show a statistically significant difference in the increase of grades between male and female students. We find that 2 of the 11 aggregation-method-and-assessment-style pairs show a benefit to male students and 1 of the 11 aggregation-method-and-assessment-style pairs shows a benefit to female students. We find all of the effect sizes to be less than 0.2, resulting in a negligible effect size. These are also shown in Table \ref{tab:demographic_results}.

Examining the data split by race, we find similar results (Fig. \ref{fig:demographics_overall}). We still find that dropping the lowest midterm exam score results in the largest grade increase. For all assessment styles and for all alternative aggregated methods, we find that A/w students outperform B/H/M/N students.

The differential impacts of the alternative midterm exam aggregation methods are shown in Table \ref{tab:demographic_results}. We find that none of the differential impacts are statistically significant, meaning that there are no statistical differences between Asian and white students and Black, Hispanic, Multiracial, and Native American students. The effects sizes, also shown in Table \ref{tab:demographic_results}, are negligible, further suggesting minimal differential impacts.

\section{Discussion}\label{sec:discussion}
Here, we provide answers to our research questions and discuss them individually.

\subsection{How do student grades vary across midterm exams within an introductory physics course?}
We find that student grades tend to vary across midterm exams within an introductory physics course. Overall, the most common midterm exam grade transition for students was to earn a lower grade on their subsequent midterm exam while the most common quintile transition was for students to remain in the same quintile on the subsequent midterm exam. However, the majority of students moved to a higher or lower quintile with about equal fraction of students moving in each direction. The distribution of transitions was also much more uniform, with the most common transition overall, staying in the same quintile, only occurring 3 percentage points more than the least common overall, moving to a lower quintile. These results largely align with Suresh and Heckler's finding of significant within-student changes in relative grade standing between midterm exams \cite{suresh_evolution_2023}.

Of the students who did not earn the same letter grade on the subsequent midterm exam, we found that the most common result was to change by one grade step compared to the previous midterm exam. Such results largely align with other studies that examine how student grades on high-stakes exams evolve over a semester \cite{ramanathan_can_2017, guidry_can_2020}.

Overall, our results suggest that while a student's grades across midterm exams are related, they are not necessarily deterministic: students who earn lower grades early on can earn a higher grade on the next midterm exam, and vice versa. However, we note that large improvements in grades are relatively rare and that most students who perform poorly on the first midterm exam do not become higher performers on the next midterm exam. From an instructional standpoint, this suggests that most students will not drastically improve their midterm exam grades on their own. As these are actual grades that already take into account existing resources offered by the instructor, supporting students to drastically improve their midterm exam grades will require resources beyond those already offered in the course.

\subsection{What impact do alternative aggregation methods have on students' overall midterm exam grades as compared to weighting each midterm exam equally?}
Unsurprisingly, we find that all three alternative midterm exam aggregation methods we examined could increase student midterm exam scores relative to the equal-weight aggregation method used in each course section of our study. 

In general, we found that dropping the lowest midterm exam score resulted in the largest grade increase, while replacing the lowest midterm exam score with the final exam score (if higher) or weighting each midterm exam differently resulted in similar grade increases to each other. In terms of magnitude, the minimum grade increase we observed was 3.5 percentage points, which occurred in the replacing the lowest midterm exam score with the final in the 5 midterm exam and a final format condition, and the largest grade increase we observed was 5.5 percentage points, which occurred in the drop the lowest midterm exam in the 3 midterm exam and a final condition. Across the four assessment styles, we generally found the effect size of implementing these alternative midterm exam aggregation methods to be small.

To contextualize these percentage increases, we can examine the impact on students' final grades. As the midterm exams in these Physics I courses tended to account for 36\% of the student's final grade in the course, we estimate that implementing these practices could raise final grades by between 1.25 to 1.98 points, potentially enough to raise students up a letter grade (for example, a 73.5 C+ could be raised to an B-  using the grading scale of this course). For courses where midterm exams constitute a larger percentage of the final grade, the effect would be even greater. 

While such a grade increase may seem trivial, there can be potential real-world consequences. For example, a student earning a B- instead of a C+ would earn an additional 0.4 grade points, which could potentially matter for students on the edge of scholarship eligibility or maintaining grade point average minimums. Further, recent work suggests students who score just above a grade cutoff in an introductory course are slightly more likely to major in that subject than students who do not score just above a grade cutoff \cite{li_grades_2024}, suggesting that small changes in grading policy could influence enrollment and decisions around a student's major.

Of course, these results rest on the assumption that student behavior would be unchanged if any of these alternative midterm exam aggregation methods were implemented in the course. If student behavior did change, then the grade increases could be different. For example, if students know their lowest midterm exam grade is dropped, they may not study for the last midterm exam if they are already happy with their aggregated grade, resulting in a lower grade increase than found here and potentially less learning, a concern frequently raised by colleagues in preliminary discussions of the results of this paper. 

While we cannot directly address this concern with our data, prior work outside of physics suggests that this may not happen in practice. For example, one study found that the majority of students do not engage in any sort of ``strategic behavior'' when midterm exam scores are dropped \cite{hadsell_grade_2010} and another found that telling students their lowest midterm exam score would be dropped did not influence their decisions to miss or not try their best on an midterm exam \cite{sewell_grade_2004}. The effect of dropping midterm exams grades on final exam scores, assumed to be a measure of overall learning in the course, is less clear as some studies have found students scored better on final exams when midterm exam scores were dropped \cite{macdermott_impact_2013, dabbour_motivating_2021} while other studies have found lower final exam grades \cite{sewell_grade_2004}.

Alternatively, if students are aware that improved performance on a later midterm exam or the final exam could mitigate the impact of an earlier bad grade, perhaps students would work harder than they otherwise would. Evidence in support of this hypothesis is mixed as some studies have found that students near grade boundaries perform better than students not near grade boundaries \cite{oettinger_effect_2002,main_impact_2014}, while others have found no difference \cite{grant_grades_2013}. Other studies suggest that students who have low grades near the middle of the term do not focus on studying for final exams as a way to improve their final grade, but rather focus on the lower-stakes aspects of the course, such as participation and in-class activities \cite{gray_effect_2022}. Whether this would happen in practice in physics courses and courses with multiple midterms, such as all of the courses studied here, is again an open question. 

Finally, while we have shown that the three alternative midterm exam aggregation methods do increase average midterm exam scores, we make no recommendation as to whether increasing grades is necessarily an ideal outcome or should be the ideal outcome. Debates around grade inflation in post-secondary education are hardly new \cite{kuh_unraveling_1999,rojstaczer_grading_2010, jephcote_grade_2021, evrard_how_2021} as are debates about curving exam grades \cite{kulick_impact_2008, thulasidas_statistical_2021, bowen_grading_2022}. Our goal for this paper is not to necessarily argue that all students should receive higher grades in their introductory physics courses, but rather to call attention to how course management decisions unrelated to student learning can have a potentially meaningful impact on the grades students earn.

\subsection{How do the impacts of the alternative aggregation methods vary based on the students' final grade?}
We found that both students who earned ``A''s and ``B''s or lower in Physics I could benefit from alternative midterm exam aggregation policies. We found that in most of the course assessment styles we examined, students earned a Physics I grade of ``B'' or lower would differentially benefit from these policies. Upon dropping the lowest midterm exam grade or variably weighting the midterm exam grades, students earned a ``B'' or lower received an additional grade bump of at least one percentage point compared to students who did earn ``A''s. As these policies are often intended to help students who may be struggling in the course, this is perhaps not surprising. Similar results have been observed before including one study that found students with the highest and lowest scores did not significantly benefit from grade dropping \cite{dabbour_motivating_2021}. While students who earned a ``B'' or lower would benefit more from any of the alternative midterm exam aggregation methods than students who did earn ``A''s, students who earn ``A'' still significantly outperform the other students, suggesting that implementing any of these methods would not hamper an instructor's ability to identify top-performing students.

\subsection{How do the impacts of the alternative aggregation methods vary based on the students’ sex or race?}
We did not find a consistent effect where any of these methods benefited female students more than male students or vice versa. Likewise, we did not find a consistent effect where any of these techniques benefited Black, Hispanic, Multiracial, or Native American more than Asian and white students or vice versa. In general, we find a result that is consistent with minoritized student groups receiving a grade boost that is not different from the grade boost received by the majoritized group. As a result, the alternative midterm exam aggregation methods explored here are equitable under the equity of fairness approach as all groups receive approximately the same benefit. However, as all groups receive approximately the same benefit, pre-existing equity gaps will be largely unaffected.

\section{Limitations and future work}\label{sec:limit_future}
Our study has multiple limitations that might affect the generalizability and applicability of our results. We discuss them here as well as future work that could resolve some of these concerns.

First, this study was a thought experiment that assumed student behavior would be unchanged if these alternative midterm exam aggregation methods were implemented in class. If students did change their behavior in response, the effects reported here could be larger or smaller depending on whether students became more or less motivated to perform well on their midterm exams.

Future work could address this limitation by collecting midterm exam grades from courses where these alternative midterm exam aggregation methods are implemented, effectively accounting for any changes to student behavior. By recalculating the grades without the alternative midterm exam aggregation method, we could obtain an estimate of how much these methods impact student grades.

Alternatively, for standardized courses with common midterm exams, we could compare aggregated student midterm exam grades in sections that implemented an alternative midterm exam aggregation method to sections that did not. Such an approach would also allow us to account for student behavior.

Second, the data for this study came from a subset of students taking Physics I rather than all students in the course. As a result, this study may not accurately represent the students who took Physics I if the students who are choosing to enroll in ECoach are fundamentally different from those in the course. As students choose to enroll in ECoach early in the semester before taking any midterm exams, we would not expect earned grades to influence enrollment in ECoach. However, students who expect to do exceptionally well in the course may choose not to use ECoach because they do not believe they need the extra help, resulting in students with higher grades potentially missing from the data. More than 90\% of Physics I students in sections that offered ECoach did enroll in ECoach and were included in this study, suggesting that any impact of missing students would likely be minimal.

To account for this, future work could collect midterm exam data directly from the gradebook so that all students can be included in the data collection. We note that whether this is possible will likely depend on institutional policies around waiving informed consent on accessing student records. Otherwise, if informed consent is required, researchers may encounter the same limitation as here and be unable to obtain a 100\% consent rate.

Finally, this study occurred at a single institution and the results may not be generalizable to other universities with different student populations or with different grading scales in Physics I. For example, while the midterm exams are still considered ``high-stakes'' at this university, the midterm exams and the final exam only constitute slightly more than half of the student's final grade in the courses studied here. For courses that have a significantly higher fraction of final grade determined by midterm exam scores (e.g., the course studied in Simmons and Heckler where the exam components constituted 70\% of the final grade \cite{simmons_grades_2020}), students' behaviors may also change in ways different from how they might change if the lowest grade were dropped for example. That is, if midterm exams make up a significant portion of their final grade, students may study more for the midterm exams as opposed to if they only make up about half of the final grade or even less.

Future work could conduct studies similar to this one at a variety of institutions and in a variety of physics departments to understand how local influences may affect the results. While we have examined four different examinations styles here, we have left the weight of aggregated midterm exam score in the final grade largely untouched.

More generally, future work could also expand beyond the grades themselves to focus on student behavior. For example, future work could interview students in courses implementing alternative midterm exam aggregation methods to understand how student behavior might be affected. For example, alternative midterm exam aggregation methods might impact student motivation, stress, and exam anxiety, the last of which is known to have a gendered impact on exam performance in introductory physics \cite{malespina_bioscience_2024}.

Further, for students who do significantly improve their midterm exam grades over time, future work could examine the strategies those students are using in comparison to strategies employed by students whose grades do not improve over time. Results from such a study could be helpful for developing interventions for students in danger of failing a course and could potentially be implemented in a platform such as ECoach, allowing the interventions to be delivered at scale to students. 

\section{Conclusion and Implications}\label{sec:conclusion}
In this paper, we have shown that student midterm exam grades in Physics I do exhibit substantial variation over time with students typically increasing or decreasing their midterm exam grades by at least one letter grade.  We then showed that alternative midterm exam aggregation methods such as dropping the lowest midterm exam score, replacing the lowest midterm exam score with the final exam score if the student performed better, and weighting the highest midterm exam score more than the lowest midterm exam score in the final grade are all effective methods for increasing the average aggregated midterm exam grade in a course, with dropping the lowest midterm exam score resulting in the largest midterm exam increase. Depending on the weight of midterm exams in the final grade, these modifications could increase students' final course grades by almost two percentage points.

We have also examined these alternative midterm exam aggregation methods as potential methods to address equity gaps. We find no evidence to support that these methods could reduce the size of equity gaps, as male and female students saw approximately the same increase in aggregated midterm exam scores as did Asian and white students compared to Black, Hispanic, Multiracial, and Native American students.

Taken together, this work finds that the alternative midterm exam aggregation methods explored here can be effective for raising average course grades, and students who earn lower grades especially benefit. However, if an instructor is hoping to reduce equity gaps in their course, we do not recommend implementing the alternative midterm exam aggregation methods explored here. At the same time, there does not appear to be any harm in doing so.

\section*{Acknowledgments}
We thank Erin Murray for her assistance in securing the data for this project. Eric F. Bell was partly supported by the National Science Foundation through grant
NSF-AST 2007065. No additional external or internal funding was received for this study. The funders had no role in study design, data collection and analysis, decision to publish, or preparation of the manuscript.

\section*{Data availability}
The data used in this study includes information protected under the Family Education Rights and Privacy Act (FERPA) and cannot be publicly shared. As the authors do not own the data and access was provided to them by the University of Michigan, access is contingent upon approval of the university. Requests should be sent to the University of Michigan's Center for Academic Innovation at ai-data-requests@umich.edu.

\bibliography{main}

\end{document}